\documentclass{aa}
\usepackage{natbib}
\usepackage{graphicx}
\usepackage{multirow}
\usepackage{xcolor}
\usepackage{txfonts}

\title{Evidence for the alignment of quasar radio polarizations \newline with large quasar group axes}
\titlerunning{Quasar radio polarization align with LQG axes}

\author{V. Pelgrims
          \inst{1}
          \and
          D. Hutsem{\'e}kers
          \inst{1}\fnmsep\inst{2}
          }

   \institute{IFPA, AGO Dept., University of Li{\`e}ge, B4000 Li{\`e}ge, Belgium\\
              \email{pelgrims@astro.ulg.ac.be}
         \and
             AEOS, AGO Dept., University of Li{\`e}ge, B4000 Li{\`e}ge, Belgium\\
             \email{hutsemekers@astro.ulg.ac.be}
             }

\begin{document}

\date{Received 16 July 2015 / Accepted April 2016}

\label{firstpage}

\abstract
{
Recently, evidence has been presented for the polarization vectors from
quasars to preferentially align with the axes of the large quasar groups (LQG)
to which they belong.
This report was based on observations made at optical wavelengths for two
large quasar groups at redshift $\sim 1.3$. The correlation suggests that
the spin axes of quasars preferentially align with their surrounding large-scale
structure that is assumed to be traced by the LQGs.
Here, we consider a large sample of LQGs built from the Sloan Digital Sky
Survey DR7 quasar catalogue in the redshift range $1.0-1.8$. For quasars
embedded in this sample, we collected radio polarization measurements
with the goal to study possible correlations between quasar polarization
vectors and the major axis of their host LQGs.
Assuming the radio polarization vector is perpendicular to the quasar spin axis,
we found that the quasar spin axis is preferentially parallel to the LQG major
axis inside LQGs that have at least $20$ members.
This result independently supports the observations at optical wavelengths.
We additionally found that when the richness of an LQG decreases,
the quasar spin axis becomes preferentially perpendicular to the LQG major
axis and that no correlation is detected for quasar groups with fewer than $10$
members.
}

\keywords{large-scale structure of Universe -- polarization -- galaxies:active -- quasars: general -- radio continuum: general}

\maketitle

\section{Introduction}
\label{sec:Intro}
The co-evolution of the spins of galaxies with their surrounding
cosmic web has been theoretically established for some time
(e.g. \citealt{White1984}; \citealt{Heavens-Peacock1988};
\citealt{Catelan-Theuns1996}; \citealt{Lee-Pen2000} and
\citealt{Hirata-Seljak2004}; see \citealt{Joachimi-et-al2015} for a
recent review). It is predicted that the spin of the dark-matter halo
as well as the spin of the central supermassive black hole (SMBH) of
a galaxy do not point in random directions of space, but
instead point towards particular directions that are determined by the
geometry of the neighbouring cosmic web
(e.g. \citealt{Aragon-Calvo-et-al2007};
\citealt{Codis-et-al2012}; see \citealt{Kiessling-et-al2015} for a
recent review).
These predictions have been supported by numerous observations
(e.g. \citealt{West1994}; \citealt{Pen-Lee-Seljak2000},
\citealt{Lee-Pen2001}, \citealt{Faltenbacher-et-al2009},
\citealt{Jones-et-al2010}; \citealt{Li-et-al2013},
\citealt{Tempel-Libeskind2013}, \citealt{Zhang-et-al2013}; see
\citealt{Kirk-et-al2015} for a recent review). Unfortunately, relying
on the apparent shapes of the galaxies that are used as a proxy of their spin
axes, these studies are limited to the low redshift ($z < 1$) Universe
because the sources need to be resolved to assess their orientations with
respect to their environment.

However, \citet{Hutsemekers-et-al2014} showed that the orientation of
the optical polarization vectors of quasars is correlated to the
orientation of the large quasar groups (LQG) to which they belong,
at redshift $z \sim 1.3$. This analysis was carried out within two large
quasar groups called the CCLQG (with 34 members) and the
Huge-LQG (with 73 members) identified by
\citet{Clowes-et-al2013} and references therein.
The authors interpreted their observations as resulting from the
alignment of the spin axes of the quasars with the orientation of the
large-scale structure to which they belong, which is assumed to be traced
by the large quasar groups.

While these alignments take place over large scales ($\geq 100\,
{h}^{-1}\rm{Mpc}$), they may reflect the recognized co-evolution
of the orientations of the spins of galaxies with the properties of their
surrounding large-scale structures.
The study of the polarization of quasars could then constitute an
additional probe of the co-evolution discussed above because it does not
suffer from the observational constraints inherent to studies relying on
the apparent morphology of galaxies \citep{Kirk-et-al2015}.
Moreover, studies involving quasars can be made at high redshift.
Therefore, it is important to confirm the correlations that involve the
polarization position angles of quasars and the characteristics
of their large-scale environments, traced here by the large quasar
groups.
To this end, instead of measuring the polarization of all quasars belonging
to a given LQG, we collect polarization measurements of quasars that
belong to a sample of LQGs and compare their polarization vectors
to the orientations of the groups to which the quasars belong.

\section{Data samples and premises}
\label{sec:DataSample}
The CCLQG and the Huge-LQG have first been identified with a
hierarchical clustering method in the quasar catalogue of the Sloan
Digital Sky Survey (SDSS) Data Release 7 (DR7). Their detection is
supported by spatial coincidence with Mg II absorbers
\citep{Clowes-et-al2013} and with a temperature anomaly in the cosmic
microwave background (\citealt{EneaR-Cornejo-Campusano2015}).
These large quasar groups have been independently confirmed
(\citealt{Nadathur2013}; \citealt{Einasto-et-al2014} and \citealt{Park-et-al2015})
using other friends-of-friends algorithms \citep[e.g.][]{Huchra-Geller1982}.
In particular, \citet{Einasto-et-al2014}
used a reliable subset of the SDSS DR7 quasar catalogue to perform
their analysis.
Their sample is defined in the redshift range of
$z \in \left[1.0,\, 1.8\right]$, in the window of the sky determined  by
$\lambda_{SDSS} \in [-55^\circ,\, 55^\circ]$ and
$\eta_{SDSS} \in [-32^\circ,\, 33^\circ],$ where $\lambda_{SDSS}$ and
$\eta_{SDSS}$ are the SDSS latitude and longitude,
respectively\footnote{https://www.sdss3.org/dr8/algorithms/surveycoords.php},
and with an additional cut in $i$--magnitude, $i \leq 19.1$.
For this sample of $22\,381$ quasars,
which we call the Einasto sample, they produced publicly
available\footnote{http://cdsarc.u-strasbg.fr/viz-bin/qcat?J/A+A/568/A46}
catalogues of LQGs that are found with a friends-of-friends
algorithm using different values of the linking length ($LL$).

We used the sample of large quasar groups built by
\citet{Einasto-et-al2014}, focusing on those groups defined by
$LL = 70\, {h}^{-1}\rm{Mpc}$. This choice is motivated by two different
reasons.
First, in \citet{Hutsemekers-et-al2014}, the alignment of
quasar morphological axes with the large-scale structures was found
in the Huge-LQG and the CCLQG. These two groups are retrieved in
the Einasto sample by using a friends-of-friends algorithm with that
$LL$ value.
Second, the richness (the number of members) of the LQGs has to be
sufficiently high to allow reliable determination of their
geometrical properties. For $LL$ values below
$70\, {h}^{-1}\rm{Mpc}$, there are at most a few LQGs with
a richness above $10$ and none above $20$. For $LL = 70\,
{h}^{-1}\rm{Mpc}$ there are several tens of rich LQGs. Above that
$LL$ value, the percolation process occurs (\citealt{Nadathur2013},
\citealt{Einasto-et-al2014}). The large quasar groups stop to grow
by including neighbouring sources and instead merge among themselves.
The number of independent rich large quasar groups thus starts to decrease
rapidly for $LL \ga 75\, {h}^{-1}\rm{Mpc}$.

We searched for polarization measurements of quasars that belong to
the Einasto sample to compare the polarization position angles
to the position angles of the groups. At optical wavelengths, there are
unfortunately too few LQG members with polarization measurements
in the compilation of \citet{Hutsemekers-et-al2005}.
Since there is a correlation between the orientation of the radio
polarization vector and the axis of the system similar to what occurs
at optical wavelengths \citep{Rusk-Seaquist1985}, we decided to consider
quasar polarization measurements from the JVAS/CLASS 8.4-GHz surveys
compiled by \citet{Jackson-et-al2007}, adopting their quality criterion on the
polarized flux ($\geq 1\, \rm{mJy}$). The choice of this sample is further
motivated below.

For the Einasto sample, we therefore searched for JVAS/CLASS radio
polarization measurements of quasars with a search radius of $0 \farcs 5$.
As in \citet{Pelgrims-Hutsemekers2015}, we constrained our sample
even more by only retaining polarization measurements if the condition
$\sigma_\psi \leq 14^\circ$ was satisfied, where $\sigma_\psi$ is the error
on the position angle of the polarization vector.
After verifying the reliability of the identifications, $185$ objects
were found. For these $185$ sources, the median of $\sigma_\psi$ is
$1.7^\circ$. With $LL = 70\, {h}^{-1} \rm{Mpc}$, $30$ of the $185$ quasars
are found to be isolated sources and $155$ belong to quasar groups
with richness $m\geq2$. To determine meaningful morphological position
angles for the LQGs, we considered at least five members as
necessary. The $86$ quasars belonging to the $83$ independent LQGs
with richness $m\geq 5$ constitute our core sample in which we investigate the
possible correlation between the quasar polarization vectors and the LQG
orientations.

\medskip

The principal contamination source of the polarization position
angle measurements at radio wavelengths is the Faraday rotation,
which takes place in our Galaxy, but also at the source. The Faraday rotation
is undesired in our study because it smears out any intrinsic correlation of the
polarization vectors with other axes.
\citet{Jackson-et-al2007} and \citet{Joshi-et-al2007} proved the reliability
of the JVAS/CLASS 8.4-GHz surveys against  any sort of biases and also
showed that the Faraday rotation at this wavelength is negligible along
the entire path of the light, from the source to us.
Galactic Faraday rotation is discussed in more detail in
Appendix~\ref{subApp:FaradayRotation}.

The radio polarization vector from the core of a quasar is expected to
be essentially perpendicular to the (projected) spin axis of its central
engine (e.g.
\citealt{Wardle2013}; \citealt{McKinney-Tchekhovskoy-Blandford2013}).
The latter can also be traced by the radio-jet axis when it is observed
in the sub-arcsecond core of the quasar. The fact that radio polarization
and jet axis are preferentially perpendicular supports the view that
the radio polarization vectors can be used to trace the quasar spin axes.
Radio polarization vectors and radio jets are known to be essentially
perpendicular (\citealt{Rusk-Seaquist1985}; \citealt{Saikia-Salter1988};
\citealt{Pollack-Taylor-Zavala2003}; \citealt{Helmboldt-et-al2007}).
This holds for the sources contained in the JVAS/CLASS 8.4-GHz
surveys, as shown by \citet{Joshi-et-al2007}.
We verified that this is also true for the sub-sample that we use here.

\begin{figure}[h]
\begin{center}
\includegraphics[width=\linewidth]{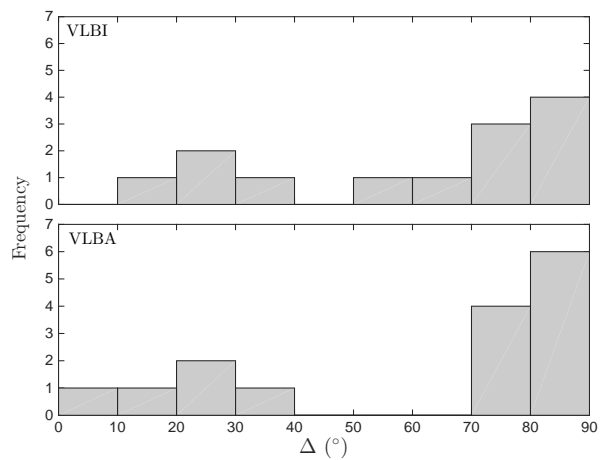}
\caption{Distributions of the acute angles between the radio polarization
vectors and the jet axes of the $13$ quasars from the VLBI compilation of
\citet{Joshi-et-al2007} (\textit{top}) and of the $15$ quasars from
the VLBA sample of \citet{Helmboldt-et-al2007} (\textit{bottom}), 
with correction for SDSS J122127.04+441129.7.}
\label{fig:DPA-vlbi-jet}
\end{center}
\end{figure}
For our sample of $41$ quasars\footnote{The cut at $m=10$ is justified
below.} in LQGs with $m\geq10$, we searched for jet axis information in
the VLBI compilation of \citet{Joshi-et-al2007} and in the VLBA sample of
\citet{Helmboldt-et-al2007}. In these catalogues, we found $13$ and $15$
sources with jet position angle measurements, respectively\footnote{These
two sub-samples are not independent. For the $9$ objects in common, the jet
position angles agree within $\sim 20^\circ$, except for one source that shows an
offset of about $72^\circ$ (SDSS J122127.04+441129.7). After
inspecting the VLBA maps \citep{Helmboldt-et-al2007}, we realized
that the sign of the position angle of the VLBA jet needs to be changed for
this object.}.
For these objects, we computed the acute angle between the polarization
vector and the jet axis.
The distribution of these angles, shown in Fig.~\ref{fig:DPA-vlbi-jet},
demonstrates that even within our small sample the radio polarizations
show a strong tendency to be perpendicular to the radio jets.
Therefore, we safely conclude that in our sample the radio polarization
vectors of the quasars trace the spin axes of the quasars and thus of
their central supermassive black holes (SMBH).
Any correlation found with the polarization vectors could then be
interpreted in terms of the quasar spin axes.

\section{Position angles of LQGs}
\label{sec:MPA_det}
To define the position angle of an LQG, we can proceed
in two ways. We can consider a group of quasars as a cloud of points on
the celestial sphere, or we can take the three-dimensional
comoving positions of the sources into account.
For either approach we determine the morphological position angle
(MPA) of an LQG through the eigenvector of the inertia tensor corresponding
to the major axis of the set of points.
\begin{figure}[b]
\begin{center}
\includegraphics[width=\linewidth]{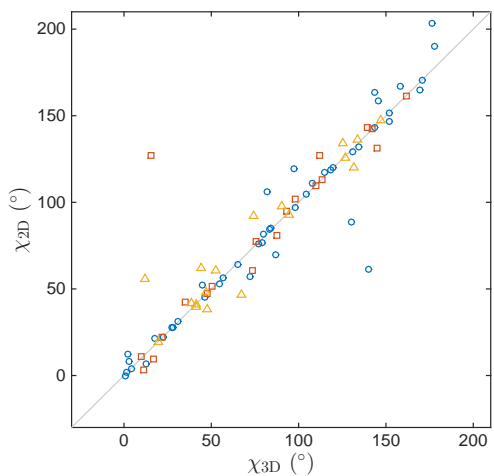}
\caption{Morphological position angles (in degrees) of the large
quasar groups determined with the two-dimensional
method ($\chi_{2\rm{D}}$) as a function of those determined with
the three-dimensional one ($\chi_{3\rm{D}}$) for the $83$ LQGs
with $m \geq 5$. Circles, squares, and triangles show LQGs
with richness $m < 10$, $10 \leq m < 20,$ and $m \geq 20$,
respectively.
Some values of $\chi_{2\rm{D}}$ have been adjusted by
$180^\circ$ for clarity.
}
\label{fig:mPAa-vs-mMPA3}
\end{center}
\end{figure}

For the two-dimensional approach, the quasar positions are projected
onto the plane tangent to the celestial sphere. The orientation
of the two-dimensional cloud of points is determined by computing its inertia
tensor, assuming quasars to be unit point-like masses. The position angle of
the eigenvector corresponding to the most elongated axis defines the
morphological position angle of the large quasar group.
We checked that this method returns position angles that are in excellent
agreement (within $1$ degree) with those obtained with an orthogonal
regression \citep{Isobe-et-al1990}.
The latter method was used in \citet{Hutsemekers-et-al2014} to define
the position angles of the quasar groups.

For the second approach, the three-dimensional comoving positions
of the quasars are used to determine the geometrical shape of an LQG
by considering its tensor of inertia, assuming quasars to be unit point-like
masses. A simple projection of the major axis of the fitted ellipsoid onto
the plane orthogonal to the line of sight defines the position angle of
the LQG.

As a result of the inclination of the system with respect to the line of sight,
the MPAs determined by the two methods may differ. In our case, we
found that they generally agree well. We show in
Fig.~\ref{fig:mPAa-vs-mMPA3} a comparison of the position angles of the
LQGs that we obtained by the two- and three-dimensional procedures.
While the two methods most often return MPAs that
agree well, these quantities can be largely different owing to the apparent
shape and the inclination of the system with respect to the line of sight. Because
the two methods return similar results, we base our discussion on the
three-dimensional approach, which is more physically motivated.
In our calculation, we assume the same cosmological model as in
\citet{Einasto-et-al2014}, that is, a flat $\Lambda$CDM Universe with
$\Omega_M = 0.27$.

For either approach, the morphological position angle of each
large quasar group is derived at the centre of mass of the group.
In general, a quasar for which we retrieved radio polarization
measurement is angularly separated from the centre of mass of its
hosting group.
Hence, the acute angle between the two orientations (the polarization
vector and the projected major axis) depends on the system of
coordinates that is used.
To overcome this coordinate dependence, we used parallel transport on the
celestial sphere to move the projected eigenvector from the centre of mass of
the group to the location of the quasar with polarization data.
By introducing $\psi$ for the polarization position angle of a
quasar and $\chi$ for the (parallel-transported) position angle of the
LQG to which it belongs, we compute the acute angle between the two orientations
as
\begin{equation}
\Delta_{\psi\chi}=90^\circ - | 90^\circ - |\psi - \chi| | \, \rm{.}
\label{eq:DeltaPsiChi}
\end{equation}
The use of the parallel transport before evaluating the acute
angle leads to coordinate-independent statistics.
Both $\psi$ and $\chi$ are defined in the range $0^\circ - 180^\circ$
and are computed in the east-of-north convention.

\section{Correlation between polarization and LQG position angles}
\label{sec:EVPA-MPAcorrelation}
\begin{figure}
\begin{center}
\includegraphics[width=\linewidth]{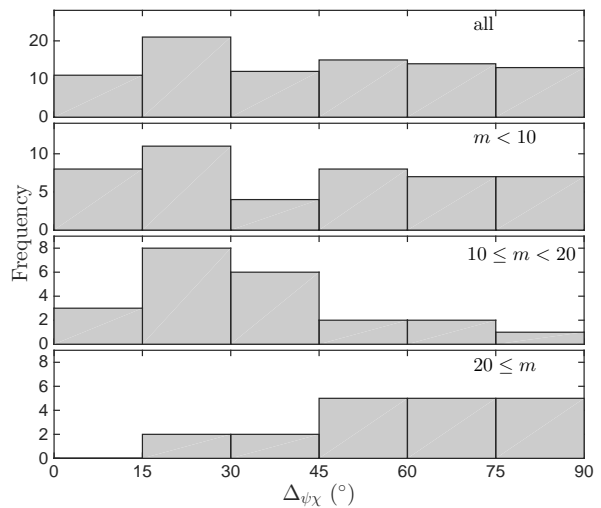}
\caption{Histogram of the distribution of $\Delta_{\psi\chi}$ (in degree)
for the $86$ quasars with polarization--LQG position-angle measurements
(top) and for the three sub-samples with richness $m < 10$,
$10 \leq m < 20,$ and $m \geq 20$.}
\label{fig:histDPA_3Dm}
\end{center}
\end{figure}
In Fig.~\ref{fig:histDPA_3Dm} (top) we show the distribution of $\Delta_{\psi\chi}$
for the $86$ quasars with polarization and LQG position-angle measurements.
The distribution of the full sample (top) does not show any departure from
uniformity. The probability given by a one-sample Kolmogorov-Smirnov (KS)
test that the distribution is drawn from a uniform parent distribution
is $P_{\rm{KS}} = 88\,\%$.
Since the alignment of optical polarization vectors with LQG orientations
was found in very rich groups, and as the accuracy of
the position angle of an LQG most likely depends on its richness, we divided our
sample into three sub-samples with $m < 10$ (45 objects), $10 \leq m < 20$
(22 objects), and $m \geq 20$ (19 objects).
For the smallest LQGs ($m < 10$), the distribution of $\Delta_{\psi\chi}$
does not show any departure from uniformity. The probability given by a
one-sample KS test that the distribution is drawn from a uniform parent
distribution is $P_{\rm{KS}}= 54 \,\%$.
However, for the larger groups, a dichotomy is observed between the
two sub-samples.
The polarization vector of a quasar belonging to a very rich LQG ($m\geq20$)
appears preferentially perpendicular to the projected major axis of the group
($\Delta_{\psi\chi}>45^\circ$), whereas the polarization vector of a quasar
belonging to an LQG with medium richness ($10 \leq m < 20$) seems to
be preferentially parallel ($\Delta_{\psi\chi} < 45^\circ$).
A two-sample KS test tells us that the probability that the two parts of the
sample with $10 \leq m < 20$ and $m \geq 20$ have their distributions of
$\Delta_{\psi\chi}$ drawn from the same parent distribution is $0.05\,\%$.

For the $19$ data points of the sub-sample of LQGs with $m \geq 20$,
$15$ show $\Delta_{\psi\chi} > 45^\circ$. The cumulative binomial
probability of obtaining $15$ or more data points with
$\Delta_{\psi\chi} > 45^\circ$ by chance is $P_{bin} = 0.96\,\%$. Of
the $22$ data points of the sub-sample of LQGs with $10 \leq m < 20$,
$17$ show $\Delta_{\psi\chi} < 45^\circ$, which gives
the cumulative binomial probability $P_{bin} = 0.85\,\%$.
These results indicate a correlation between the
position angle of the major axis of an LQG and the radio polarization
vector of its members in rich ($m\geq10$) quasar groups.

\begin{figure}
\begin{center}
\includegraphics[width=\linewidth]{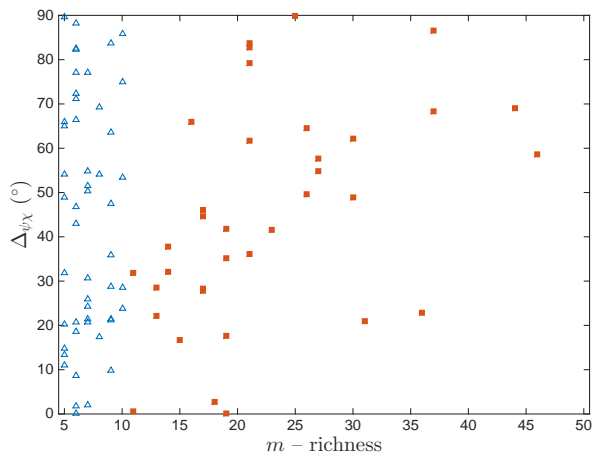}
\caption{$\Delta_{\psi\chi}$ versus the richness $m$ of the LQGs. For
$m\leq10$ and $m\geq 11$, symbols are triangles and filled squares,
respectively.}
\label{fig:DPA3D-vs-m}
\end{center}
\end{figure}

The dichotomy between the two sub-samples of LQGs with $10\leq m < 20$
and $m\geq20$ is also illustrated in Fig.~\ref{fig:DPA3D-vs-m}, where we plot
the $\Delta_{\psi\chi}$ of each quasar against the richness of its hosting LQG.
Surprisingly, for $m \geq 11$, we even see a possible linear correlation
of $\Delta_{\psi\chi}$ with the richness of the large quasar groups.
A Spearman correlation test on the pairs $\Delta_{\psi\chi} - m$ gives
a rank-order correlation coefficient of $0.54$ with a probability of
obtaining this result by chance of $0.08\, \%$.  
For $m < 10$, there is no specific trend of $\Delta_{\psi\chi}$ with
the richness in agreement with the distribution seen in
Fig.~\ref{fig:histDPA_3Dm}. To understand whether this
lack of correlation for $m<10$ is due to a larger uncertainty of
the major axis position angles of the LQGs, as might naively
be expected
for the smallest groups, we estimated the confidence interval of the
morphological position angle using the bootstrap method described in
Appendix~\ref{subApp:MPAs-and-err}.
Keeping only MPAs for which the half-width confidence interval is below
$20^\circ$ ($27$ objects out of $45$), the distribution of $\Delta_{\psi\chi}$
remains uniform with $P_{\rm{KS}} = 72\,\%$. The absence of alignments
in large quasar groups with richness $m < 10$ is therefore likely
to be real.
On the other hand, the uncertainties, both on the position
angles of the major axes of the LQGs and on the polarization
position angles, cannot account for the correlations that we
report. The introduction of poorly defined orientations in our
analysis can only scramble an existing correlation. The same argument
applies to the contamination of the polarization position angles by
Galactic Faraday rotation, which is found to be negligible (see
Appendix~\ref{subApp:FaradayRotation}).

\medskip

In summary, our analysis shows that the quasars that belong to very
rich ($m\geq20$) large quasar groups have polarization vectors
preferentially perpendicular to the projected major axes of their hosting
LQGs.
The polarization vectors then become more often parallel to the LQG
axes when $m$ decreases before no correlation is observed for the
smallest ones ($m<10$).

\section{Discussion}
\label{sec:Interpretation}
As discussed in Sect.~\ref{sec:DataSample}, the radio polarization vector
of a quasar is expected to be essentially perpendicular to the spin axis of
its central engine.
The correlation that we found between the polarization vectors and the
major axes of the host large quasar groups might thus reflect an existing
link between the spin axes of the quasars and the major axes of the host
LQGs.

Our analysis independently supports the view that the spin axis of
quasars that belong to very rich LQGs are preferentially parallel to the
major axes of their hosting LQGs, as found in
\citet{Hutsemekers-et-al2014}. In addition, we found that the
quasar-spin axes become preferentially perpendicular to the LQG's
major axes as the richness of the LQGs decreases, down to $m\geq10$.

Regardless of the richness dependence that we discuss below, our
observation also suggests that the quasar spin axes have an intrinsic
tendency to align themselves within their host large quasar groups. The
observations of \citet{Jagannathan-Taylor2014} that radio jets in the GRMT
ELAIS N1 Deep Field align with each other over scales of
$50 - 75 \,{h}^{-1}\,\rm{Mpc}$ at redshift $z \ga 1$ support our interpretation
\citep{Taylor-Jagannathan2016}.

\medskip

As discussed in the introduction, the fact that the spin axes of black
holes are found to align with their surrounding large-scale
structures, which are assumed to be traced by LQGs, could be understood in the
framework of the tidal torque theory if we accept that these predictions
can be extrapolated to larger scales.
However, and as far as we
know, a richness dependence of the relative orientation is not a
predicted feature of this theory. We therefore explore our data
set to determine whether the richness dependence hides another
dependence that could be more physically motivated.

For this exploratory analysis, we relied on our core sample of LQGs with
$m \geq 10$ and for which we have radio polarization measurements for
at least one of their quasar member (Table~\ref{tab:dataTable}).
For LQGs with $m\geq 10$, the median of the diameters (the largest
separation between two members in groups) is about
$310\,{h}^{-1}\,\rm{Mpc}$ and the median separation of the closest
pairs is about $45\,{h}^{-1}\,\rm{Mpc}$.

\subsubsection{Richness dependence and quasar intrinsic properties}
N-body simulations have shown that the direction
towards which the spin axes of dark-matter haloes preferentially point
relative to the large-scale structure depends on their mass
(\citealt{Codis-et-al2015}).
As the masses of the SMBH and the host dark-matter halo are thought
to be linked and as their spin axes might be aligned at high redshift
(\citealt{Dubois-Volonteri-Silk2014}), we searched for a possible
dependence of $\Delta_{\psi \chi}$ with the black hole mass.
Using the Spearman correlation test, we found that the SMBH
masses (from \citealt{Shen-et-al2011}) do not show any correlation with $\Delta_{\psi \chi}$ or with the richness of the host LQG.
The same conclusion is reached if we consider other quasar properties
reported in \citet{Shen-et-al2011}, such as the radio-loudness, the width
of the emission lines, or the redshift. There is thus apparently no hidden
relation of $\Delta_{\psi \chi}$ with quasar properties that could explain
the richness dependence of $\Delta_{\psi \chi}$.

\subsubsection{Richness dependence and LQG characteristics}
In Sect.~\ref{sec:MPA_det} we used the inertia tensors of the large
quasar groups to assign their orientations in the three-dimensional
comoving space. This resulted in fitting ellipsoids to the quasar systems.
Here, we use the relative lengths of the principal axes of the ellipsoids to
characterize their shapes. We found no correlation of $\Delta_{\psi \chi}$,
or of the richness of the groups $m$, with geometrical characteristics of
the LQGs such as their oblateness or departure from spherical or
cylindrical symmetry.

A quantity that is related to the richness of an LQG and that could
have a better physical meaning is its density. We can naively define
the density of an LQG as $\rho = m/V$, where $V$ is the comoving
volume of its fitted ellipsoid. In our core sample of LQGs, there is
a relation between the richness and the density: the richer a large quasar
group, the lower its density. We then applied a Spearman correlation
test to the pairs $\Delta_{\psi \chi} - \rho$, which resulted in
rank-order correlation coefficient of $-0.50$ with a probability of
obtaining this result by chance of $0.19 \%$, if we consider all the
LQGs with $m \geq 11$ and at least one $\Delta_{\psi \chi}$
measurement (the sub-sample studied in
Sect.~\ref{sec:EVPA-MPAcorrelation}). SMBHs spin axes are thus
parallel to the host LQG axis when the density of the latter is low
and perpendicular when the density is high. The strength of this
correlation is similar to the strength of the $\Delta_{\psi \chi} - m$
correlation. The dependence between $\Delta_{\psi \chi}$ and the
richness $m$ could then reflect a dependence between $\Delta_{\psi
\chi}$ and $\rho,$ which might be easier to interpret.
\citet{Codis-et-al2015} have shown that the spin axes of the
dark-matter haloes of galaxies are preferentially parallel to their
neighbouring filaments when the halo masses are low or,
equivalently, when the density of their cosmic environment is low.
When the density of the environment increases, the halo spin axis
starts to avoid alignment with the filaments to finally point
preferentially perpendicular to them.
Our observations might thus support these predictions if we assume
that, at least at redshift $z\geq 1.0$, (\textit{i}) a similar behaviour can
be expected for the spin axes of the central SMBH of quasars, (\textit{ii})
the density of an LQG reflects the density of the surrounding cosmic web,
(\textit{iii}) the LQGs can be used to trace the large-scale structures, and
(\textit{iv}) correlations that occur between the galaxy spin axes and filaments
also occur at larger scales between quasar-spin axes and LQG major axes.

\section{Conclusion}
Studying the optical linear polarization of quasars that belong to two
rich large quasar groups located at redshift $z \sim 1.3$,
\citet{Hutsemekers-et-al2014} noted that the quasar polarization
vectors do not point randomly with respect to the group structural
axes. Using object spectra, they furthermore inferred that the spin axis of
quasars, that is, the supermassive black hole spin axes, are
preferentially parallel to the major axes of the quasar groups. We here examined this result by
considering the sample of large quasar groups published by
\citet{Einasto-et-al2014}, which is drawn from the SDSS DR7 quasar
catalogue in the redshift range $1.0 \leq z \leq 1.8$. We used the
LQGs defined with a linking length of $70\, h^{-1}\, \rm{Mpc}$.
Because too few optical polarization data are available for the
quasars that belong to the Einasto sample, we used radio polarization
measurements from the JVAS/CLASS 8.4-GHz surveys
(\citealt{Jackson-et-al2007}). This sample is claimed to be free of biases
that might affect the polarization angles. Furthermore, the polarization
vectors measured at 8.4 GHz have a strong tendency to be perpendicular
to the spin axes of the SMBHs in the quasar cores. To compare
the position angles of the quasar polarization vectors with the position
angles of the systems to which the quasars belong, we studied the LQGs
through their inertia tensors in the three-dimensional comoving space.

For rich quasar groups ($m \geq 20$), we found that the spin axes of
the SMBHs are preferentially parallel to the major axes of their host
LQGs.
This result adds weight to the previous finding by
\citet{Hutsemekers-et-al2014} that in two large quasar groups the quasar
spin axes (inferred from optical polarizations) align with the group axes.
Combined with the initial discovery, our analysis indicates that the
alignments of the SMBH spins axes with the LQG major axes do not
depend on the quasar radio loudness.

Additionally, the use of a large sample of LQGs allowed us to probe
the alignments for a wide variety of quasar systems.  We unveiled a
surprising correlation: the relative orientations of the spin axes of
quasars with respect to the major axes of their host LQGs appear to
depend on the richness of the latter, or equivalently on the density
of objects. The spin axes of SMBHs appear preferentially parallel to
the major axes of their host LQGs when the latter are very rich (or
have a very low density), while the spin axes become preferentially
perpendicular to the LQG major axes when the richness decreases to
$m \geq 10$ or, equivalently, when the quasar density
increases to $1.5\, 10^{-5}\,(h^{-1}\, \rm{Mpc})^{-3}$. No correlation
is observed below this richness or above this density. Possible
interpretations were discussed in Sect.~\ref{sec:Interpretation}, but this
intriguing feature needs to be confirmed.

In agreement with the view that the richest large quasar
groups at high redshift most likely represent the progenitors of complexes or chains of
superclusters \citep{Einasto-et-al2014}, the correlation that we found
might be the high-redshift counterpart of the alignments at $z \sim 0$
of clusters of galaxies with the superclusters in which they are
embedded (e.g. \citealt{Einasto-Joeveer-Saar1980}; \citealt{West1999}).

\begin{acknowledgements}
We thank D. Sluse for suggestions regarding the presentation
of the work, M. Einasto for interesting discussions, and our referees
for pointing out a mistake in the first version of the paper and for
suggestions that improved the presentation of our results.
D.H. is Senior Research Associate at the F.R.S.--FNRS.
This work was supported by the Fonds de la Recherche
Scientifique -- FNRS under grant 4.4501.15. This research has made
use of the NASA/IPAC Extragalactic Database (NED) which is operated by
the Jet Propulsion Laboratory, California Institute of Technology,
under contract with the National Aeronautics and Space
Administration. This research has made use of NASA's Astrophysics Data
System.
\end{acknowledgements}

\footnotesize{
\bibliographystyle{aa}
\bibliography{mn-jour,myReferences}
}

\begin{appendix}
\label{appendix}

\section{Contamination by Galactic Faraday rotation}
\label{subApp:FaradayRotation}
Although \citet{Jackson-et-al2007} and \cite{Joshi-et-al2007}
stated that the Faraday rotation of our Galaxy can be
neglected for quasars observed at $8.4$ GHz, it is important to verify
that this contamination is indeed negligible for our sample.

We used the all-sky Galactic Faraday map produced by
\citet{Oppermann-et-al2015} to verify that the Galactic Faraday rotation
can be neglected in our analysis. From their map of rotation measures,
we extracted the whole sky window covered by the Einasto sample. For the
entire window, the distribution of the Faraday rotation angles at 8.4 GHz
that is due to the Galactic magnetic field has a mean at $0.6^\circ$ and a
standard deviation of about $1^\circ$.
For the source locations of our sample with polarization
measurements (the $185$ JVAS/CLASS sources), the contamination is
even lower with a mean of $0.5^\circ$ and a standard deviation of
$0.6^\circ$.
We conclude that the Galactic Faraday rotation can be neglected because the
rotation angles are within the error bars of the polarization data.

\section{Position angles and their uncertainties}
\label{subApp:MPAs-and-err}
To quantify the uncertainties of the morphological position
angles (MPA) that characterize the large quasar groups, we used the
bootstrap method.
This procedure allows properly accounting for the circular nature of
the data (\citealt{Fisher1993}).
For a given LQG of richness $m$, we produced $N_{\rm{sim}}$
bootstrap LQGs with the same number of members, allowing
replicates. The position angle of each bootstrap LQG is determined
through the inertia tensor procedure used for real groups.
As this procedure is not properly defined for groups resulting in only
one point, we took care in the generation of LQGs to avoid bootstrap
samples consisting of $m$ replications of the same source.
The probability that such a configuration occurs is $m^{1-m}$. Hence,
the rejection procedure can only affect poor LQGs.
We note that even for those poor LQGs, the effect of theses configurations
on the evaluation of the confidence interval is negligible ($\ll 1^\circ$).
For a given group of quasars, we therefore collected a corresponding
distribution of $N_{\rm{sim}}$ estimates of the morphological position
angle.
From this distribution, we evaluated the mean and its corresponding
confidence interval. For a distribution of axial-circular quantities $\chi_k$
such as the position angle of the LQG major axes, the mean is computed
as (\citealt{Fisher1993})
\begin{equation}
\label{eq:meanPA}
\bar{\chi} = \frac{1}{2}\, \arctan \left(\frac{\sum_{k=1}^{m} \sin 2\chi_k}{\sum_{k=1}^{m} \cos 2\chi_k} \right)\,.
\end{equation}
Since there is no proper definition of the standard deviation for axial-circular
data, we evaluated the confidence interval of the unknown mean at the
$100\,(1-\alpha)\%$ level as follows \citep{Fisher1993}. We defined
\begin{equation}
\label{eq:gamma_b}
\gamma_k = \frac{1}{2} \arctan \left( \frac{\sin \left( 2(\chi_k - \bar{\chi}) \right)}{\cos \left( 2(\chi_k - \bar{\chi}) \right)} \right)
,\end{equation}
where $k=1, ..., N_{\rm{sim}}$. The $\gamma_k$ were defined in the range $\left[-90^\circ, \, 90^\circ \right]$.
Then we sorted the $\gamma_k$ in increasing order to obtain
$\gamma_{(1)} \leq ... \leq \gamma_{(N_{\rm{sim}})}$. If $l$ is the integer part
of $\frac{1}{2} \left(N_{\rm{sim}} \alpha +1 \right)$ and $u=N_{\rm{sim}} - l$,
the confidence interval for $\bar{\chi}$ is
$\left[ \bar{\chi} + \gamma_{(l+1)}, \, \bar{\chi} + \gamma_{(u)} \right]$.
We chose to compute the confidence interval of $\bar{\chi}$ at the $68\%$ confidence level.
We defined the half-width of the confidence interval as
$\rm{HWCI} = \left( \gamma_{(u)} - \gamma_{(l+1)} \right)/2$.

\begin{figure}
\begin{center}
\includegraphics[width=\linewidth]{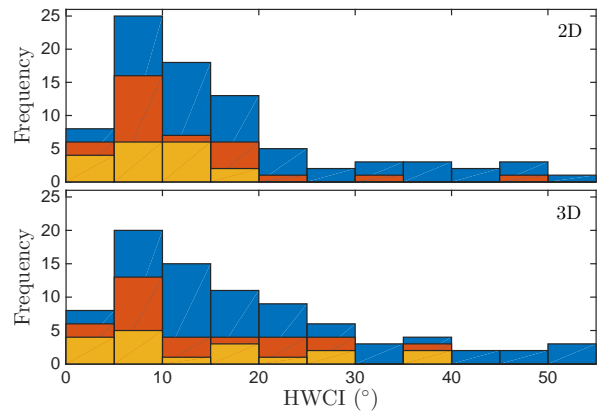}
\caption{Histograms of the half-width confidence interval (HWCI) for
the MPA values obtained for the 83 independent LQGs with the two-
(\textit{top}) and three-dimensional (\textit{bottom}) approaches
(see Sect.~\ref{sec:MPA_det}).
The HWCIs are evaluated using the bootstrap method as explained in the text.
Histograms are for $m\geq5$ (blue), $m\geq10$ (orange), and $m\geq20$ (yellow).}
\label{fig:HWCI-chi2D3D}
\end{center}
\end{figure}

Using the bootstrap method with $10\,000$ simulations, we evaluated
the HWCIs for the 83 (independent) LQGs of our sample.
The distribution of HWCI corresponding to the three-dimensional
evaluation of the morphological position angles is shown in
Fig.~\ref{fig:HWCI-chi2D3D} for different cuts in richness and is
compared to those obtained from the two-dimensional procedure.
In general, the HWCIs are lower when the two-dimensional procedure
is chosen to estimate the MPAs. The highest
values of the HWCIs ($\geq 20^\circ$) can in general mostly be
attributed to poor
LQGs with $m<10$, as naively expected.

\section{Data sample}
\label{subApp:DataSample}
\begin{figure}
\begin{center}
\includegraphics[width=\linewidth]{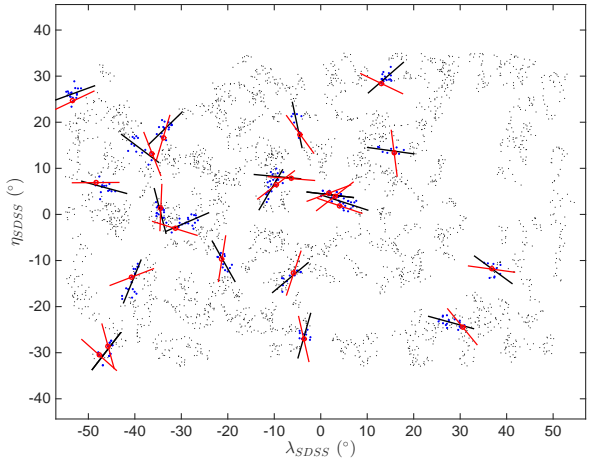}
\caption{Projection on the sky, in SDSS coordinates, of the LQGs with a
richness in the range $10 - 19$ (clouds of grey dots).
The LQGs containing at least one quasar with polarization measurements
(circled in red) are highlighted in blue. The black lines trace the orientation
of the projected major axes of the groups (here at the centres of masses).
The red lines give the orientations of the polarization vectors. All lines are
of equal lengths for clarity. The polarization vectors are preferentially
parallel to the projected major axes of the groups.}
\label{fig:skyPlot_less20}
\end{center}

\end{figure}

\begin{figure}
\begin{center}
\includegraphics[width=\linewidth]{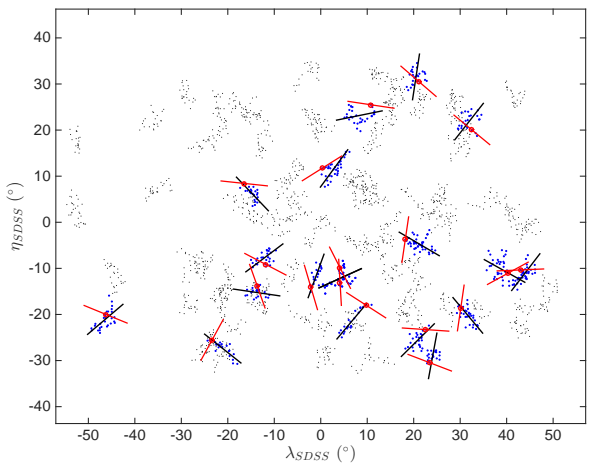}
\caption{Same as Fig.~\ref{fig:skyPlot_less20}, but for the LQGs with
at least 20 members. The polarization vectors are preferentially
perpendicular to the projected major axes of the groups.}
\label{fig:skyPlot_geq20}
\end{center}
\end{figure}

Table~\ref{tab:dataTable} summarizes the data for the $41$ quasars
hosted in LQGs with $m\geq10$.
For each quasar for which we collected polarization measurements, we
list the SDSS name, its redshift, the position angle of its polarization
vector ($\psi$), the identification index of the LQG to which it belongs
(following the numbering of \citet{Einasto-et-al2014}), the richness of the
group, and the position angle of the projected major axis ($\chi_{3\rm{D}}$).
In Figs.~\ref{fig:skyPlot_less20} and~\ref{fig:skyPlot_geq20} we show the
projection on the sky of the medium and high richness parts of the LQG
sample with $10 \leq m<20$ and $m\geq 20$,
respectively.
We highlighted those LQGs for which we retrieved the radio polarization
for at least one member and show the orientations of the projected major
axes along with the polarization vectors of the quasars.

\begin{table*}
\begin{center}
\caption{\label{tab:dataTable}Data for the $41$ quasars with
$\Delta_{\psi \chi}$ measurements and belonging to rich ($m \geq 10$)
LQGs.}

\begin{tabular}{lcr | rrr}
\hline
\hline
SDSS name & $z$ & $\psi$ ($^\circ$) & LQG ID & $m$ & $\chi_{3\rm{D}}$ ($^\circ$) \\
\hline

074809.46+300630.4      &       1.6942  &            13.3               &              7 &               18              &               15.9            \\
083740.24+245423.1 &    1.1254  &       58.0            &          184   &               15              &               74.8            \\
090910.09+012135.6      &       1.0255  &     130.4             &          410   &               10              &               36.2            \\
091204.62+083748.2      &       1.5388  &     103.3             &          356   &               30              &               41.1            \\
091439.42+351204.5      &       1.0738  &     127.6             &          267   &               14              &               95.5            \\
091641.76+024252.8      &       1.1019  &     162.0             &          410   &               10              &               35.5            \\
091648.90+385428.1      &       1.2656  &     163.7             &          396   &               17              &               12.1            \\
093105.33+141416.4      &       1.0997  &       56.2            &          548   &               17              &               10.2            \\
094148.11+272838.8      &       1.3063  &     161.5             &          652   &               19              &               143.9   \\
095956.04+244952.4      &       1.4803  &       91.1            &          844   &               19              &               49.3            \\
104552.72+062436.4      &       1.5091  &       25.7            &        1199            &               21              &               126.6   \\
104831.29+211552.2      &       1.4810  &         0.2           &         1215            &               14              &               142.3   \\
105431.89+385521.6      &       1.3662  &       82.3            &       1266            &               23              &               123.8   \\
112229.70+180526.4      &       1.0414  &     156.1             &       1437            &               21               &               94.3            \\
112814.74+225148.9      &       1.0809  &     112.7             &       1453            &               26               &               48.2            \\
113053.28+381518.6      &       1.7413  &       44.7            &       1507            &               13               &               22.7            \\
114658.29+395834.2      &       1.0882  &       89.3            &       1501            &               11               &               89.8            \\
115232.86+493938.6      &       1.0931  &     139.0             &       1643            &               10              &               162.7   \\
115518.29+193942.2      &       1.0188  &       16.0            &       1716            &               11               &               47.8            \\
120518.69+052748.4      &       1.2956  &      166.9          &       1857            &               13               &               15.5            \\
121106.69+182034.2      &       1.5150  &     163.0             &       1925            &               21               &               19.1            \\
122127.04+441129.7      &       1.3444  &       57.9            &       2002            &               36               &               35.0            \\
122847.42+370612.0      &       1.5167  &       70.7            &       1996            &               17               &               98.5            \\
123505.80+362119.3      &       1.5983  &       54.2            &       1996            &               17               &               98.8            \\
123736.42+192440.5      &       1.5334  &     178.2             &       2011            &               37               &               66.5            \\
123757.94+223430.1      &       1.4175  &     155.0             &       2011            &               37               &               68.3            \\
123954.13+341528.8      &       1.1698  &      111.6            &       2151            &               19              &               111.7           \\
130020.91+141718.5      &       1.1060  &     125.5             &       2136            &               21               &               41.7            \\
133915.90+562348.1      &       1.4254  &     115.3             &       2287            &               31               &               94.3            \\
134208.36+270930.5      &       1.1898  &       18.0            &       2650            &               44              &               129.0   \\
134821.89+433517.1      &       1.1140  &         8.4           &       2563            &               10              &               113.5           \\
135116.91+083039.8      &       1.4398  &       97.5            &       2580            &               26               &               47.9            \\
135351.58+015153.9      &       1.6089  &     111.9             &       2714            &               21               &               14.7            \\
140214.81+581746.9      &       1.2673  &     137.9             &       2809            &               16               &               71.9            \\
142251.89+070025.9      &       1.4505  &     145.0             &       2813            &               19              &               109.9   \\
142330.09+115951.2      &       1.6127  &       15.3            &       2961            &               30              &               146.4   \\
145420.85+162424.3      &       1.2763  &      110.5            &       3198            &               10              &               139.0   \\
150124.63+561949.7      &       1.4670  &      166.2            &       3126            &               27               &               43.8            \\
150910.11+161127.7      &       1.1474  &        75.3           &       3219            &               46              &               134.0   \\
152037.06+160126.6      &       1.4669  &      103.1            &       3408            &               27               &               48.4            \\
152523.55+420117.0      &       1.1946  &      164.8            &       3375            &               25               &               74.8            \\

\hline
\end{tabular}
\tablefoot{Column 1 gives the SDSS quasar name, Col. 2 the redshift $z$,
Col. 3 the polarization position angle ($\psi$) in degrees (east-of-north),
Col. 4 the identification index from the catalogue of \citet{Einasto-et-al2014} of
the LQG to which the quasar appertains, Col. 5 the number of member in that
LQG $m$ and Col. 6 the position angle ($\chi_{3\rm{D}}$, in degrees, east-of-north)
of the major axis of the LQG when projected on the sky and parallel transported at
the location of the quasar for which we collected polarization measurements.}
\end{center}
\end{table*}


\end{appendix}

\label{lastpage}
\end{document}